\documentclass[superscriptaddress,reprint,amsmath,amssymb,aps,prl]{revtex4-1}

\usepackage{txfonts}
\usepackage{graphicx}
\usepackage{dcolumn}
\usepackage{bm}
\usepackage{xcolor}
\usepackage{tabularx}
\usepackage{ulem}
\usepackage[colorlinks,linkcolor=blue,citecolor=blue,urlcolor=blue]{hyperref}

\def\bs#1{\boldsymbol{#1}}

\begin{document}

\title{Berry-curvature-induced linear magnetotransport in magnetic Weyl semimetals}

\author{Zetao \surname{Zhang}}
\affiliation{State Key Laboratory of Low Dimensional Quantum Physics and Department of Physics, Tsinghua University, Beijing 100084, China}

\author{Yizhou \surname{Liu}}
\email{yizhou.liu@weizmann.ac.il}
\affiliation{Department of Condensed Matter Physics, Weizmann Institute of Science, Rehovot 76100, Israel}

\author{Wenhui \surname{Duan}}
\affiliation{State Key Laboratory of Low Dimensional Quantum Physics and Department of Physics, Tsinghua University, Beijing 100084, China}
\affiliation{Institute for Advanced Study, Tsinghua University, Beijing 100084, China}
\affiliation{Frontier Science Center for Quantum Information, Beijing 100084, China}

\date{\today}

\begin{abstract}
Magnetotransport such as the giant magnetoresistance and Hall effect lies at the heart of fundamental physics and technologies. Recently, some experiments have clearly demonstrated linear magnetotransport (LMT) proportional to magnetic field but the underlying physical mechanism is still unclear. In this work, we show that Berry curvature effect is a new mechanism dominating the LMT. The Berry-curvature-induced LMT widely exists in 66 out of 122 magnetic point groups. For typical magnetic Weyl semimetals Co$_3$Sn$_2$S$_2$ and ferromagnetic MnBi$_2$Te$_4$, Berry curvature induces LMT conductivities reaching orders of $10^4$ and $10^2$ ${\rm \Omega^{-1}m^{-1}}$ per tesla, respectively, which are tunable through magnetization canting induced by moderate magnetic fields. We further reveal that Berry-curvature-induced LMT can be detected by Hall effect and especially intrinsic magnetoresistance exceeding $100\%$ per tesla insensitive to the sample quality. Our results agree with recent experiments and uncover the important role of Berry curvature in LMT.
\end{abstract}

\maketitle

\textit{Introduction.}---Concepts of Berry phase and curvature have revolutionized modern condensed matter physics \cite{Xiao2010review, vanderbilt2018berry}, and are essential for the description of various topological materials \cite{hasan2010colloquium, qi2011topological, Yan2017, Armitage2018WSM_review, sato2017topological}. In Weyl semimetals (WSMs), Berry curvature is especially prominent, due to the existence of two-fold degenerate points acting as monopoles of Berry curvature \cite{Yan2017, Armitage2018WSM_review}. As a consequence, Berry curvature introduces strong modifications to the electron motion \cite{sundaram1999wave, Xiao2010review} and is closely related to various exotic magnetotransport phenomena \cite{Nandy2017PHE, Gao2019review, Liu2021QIs} such as the negative magnetoresistance (MR) \cite{Son2013NMR_theory, burkov2014chiral}, a chiral magnetic effect \cite{fukushima2008chiral, Zyuzin2012CA,Armitage2018WSM_review} which originates from chiral anomaly as inspired from the studies in particle physics \cite{adler1969axial, bell1969pcac, nielsen1983adler} and has been confirmed experimentally \cite{Huang2015NMR_exp_TaAs}.

\begin{figure}
    \centering
    \includegraphics[width=\linewidth]{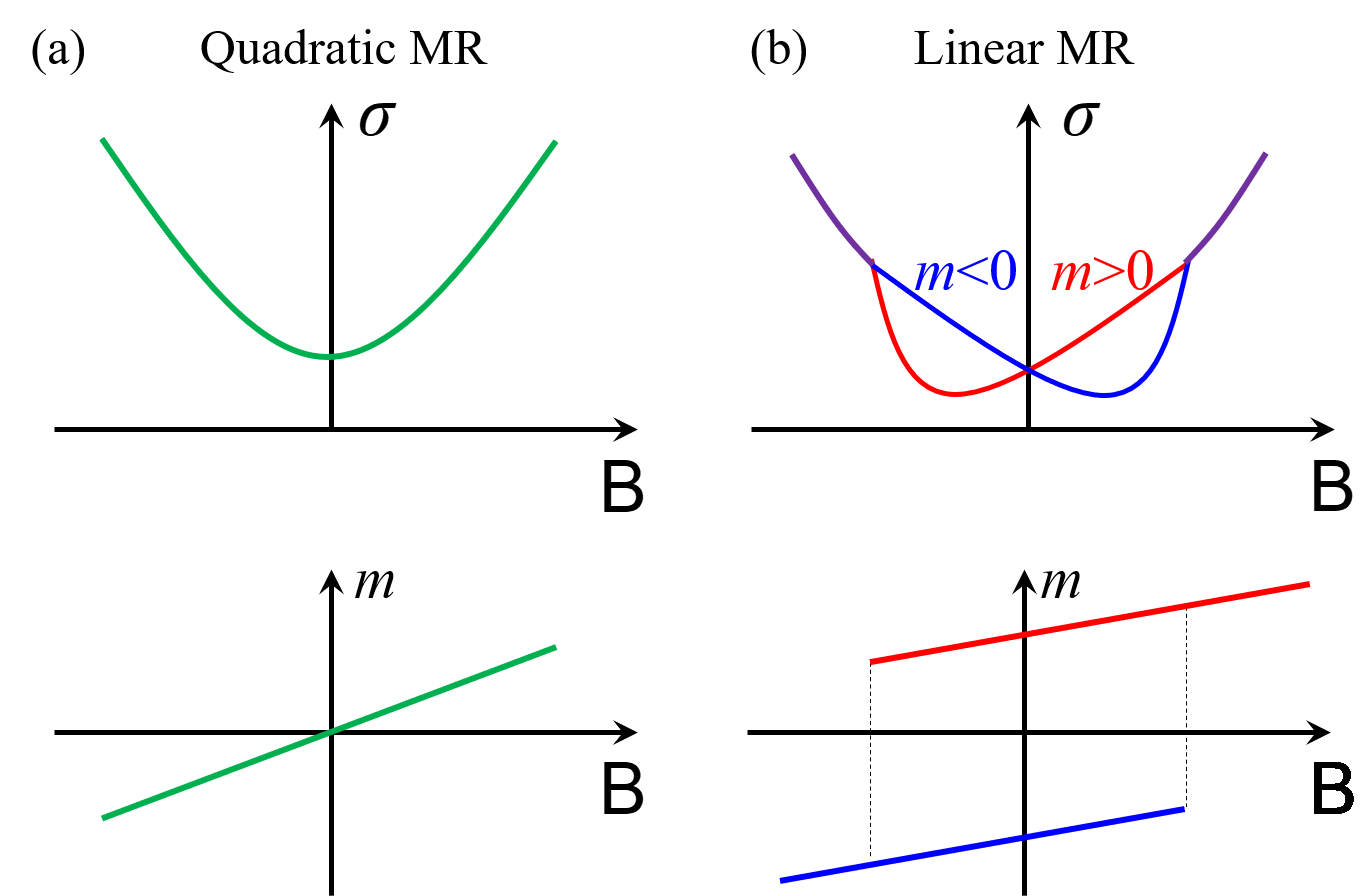}
    \caption{Schematics of (a) quardratic magnetoresistance (MR) in nonmagnetic materials and (b) linear MR in magnetic materials at small magnetic field $B$, respectively. The upper (lower) panel shows the electrical conductivity $\sigma$ (magnetization $m$) as a function of external magnetic field $B$.}
    \label{fig1}
\end{figure}

Lying at the heart of physics and technologies, magnetotransport is characterized by the Hall effect and MR corresponding to the antisymmetric and symmetric parts of resistivity tensor $\rho_{\alpha\beta}$, respectively. According to Onsager's relation: $\rho_{\alpha\beta}(\bm{m},\bm{B})=\rho_{\beta\alpha}(-\bm{m},-\bm{B})$ with $\bm{m}$ and $\bm{B}$ being the magnetization and magnetic field, respectively \cite{onsager1931reciprocal, Tokura2018review}, for nonmagnetic materials the lowest order dependence of MR on magnetic field is $B^2$, which is the case of ordinary MR effect (Fig. \ref{fig1}(a)). However, for magnetic materials linear MR proportional to $B$ could exist (Fig. \ref{fig1}(b)). In linear magnetotransport (LMT) proportional to $B$, except for the ordinary Hall effect originating from Lorenz
force, linear MR has also been observed in various topological materials, such as quantum anomalous Hall insulators \cite{chang2013experimental, deng2020quantum}, Dirac semimetals \cite{feng2015large,liang2015ultrahigh,novak2015large,narayanan2015linear,zhao2015anisotropic} and topological insulators \cite{tang2011two,wang2012room,wang2012linear}, where the underlying mechanisms are attributed to topological edge states \cite{checkelsky2012dirac,chang2013experimental}, Zeeman splitting \cite{feng2015large,liang2015ultrahigh,wang2012linear}, disorders \cite{parish2003non, parish2005classical,novak2015large,narayanan2015linear} or Landau levels \cite{abrikosov1998quantum, abrikosov2000quantum,zhao2015anisotropic,wang2012room}. Although LMT is extensively studied, the role of Berry curvature in LMT is still unclear in realistic materials.

In this work we clarify the role of Berry curvature in LMT and reveal its dominance in recent LMT experiments on magnetic WSMs. We find Berry-curvature-induced LMT is widely supported by 66 out of all 122 magnetic point groups (MPGs). Based on first-principles calculations, we predict Berry curvature induces LMT conductivities up to $10^4$ and $10^2$ ${\rm \Omega^{-1}m^{-1}}$ per tesla in representative magnetic WSMs Co$_3$Sn$_2$S$_2$ and ferromagnetic (FM) MnBi$_2$Te$_4$, respectively, which can be tuned through magnetization canting induced by moderate magnetic fields. Experimentally, Berry-curvature-induced LMT can be characterized by signatures including linear Hall resistance and especially intrinsic linear MR up to $100\%$ per tesla. Our analyses are well consistent with recent LMT experiments on Co$_3$Sn$_2$S$_2$ and FM MnBi$_2$Te$_4$ \cite{jiang2021chirality,li2021gate, lei2020surface, zhu2020negative,huan2021magnetism}, supporting that LMT in these two materials is mostly contributed by Berry curvature through modifying density of states (DOS) \cite{explainMDOS}. These results deepen understandings of Berry-curvature-induced LMT and reveal the importance of Berry curvature in LMT.

\textit{Theory of Berry-curvature-induced LMT.}---Based on semiclassical theory \cite{sundaram1999wave, Xiao2010review}, the electric current is $j_\alpha = \sigma^{0}_{\alpha\beta} E_\beta + \sigma^{\text{LMT}}_{\alpha\beta\gamma} E_\beta B_\gamma$ up to linear responses of both electric and magnetic fields $\bm{E}$ and $\bm{B}$ (see details in section S1 of Supplemental Material (SM) \cite{SM}), where $\sigma^{0}_{\alpha\beta}$ includes magnetic-field-independent Drude and anomalous Hall conductivities and $\sigma^{\text{LMT}}_{\alpha\beta\gamma}B_\gamma$ gives Berry-curvature-induced LMT conductivity. $\sigma^{\text{LMT}}_{\alpha\beta\gamma}$ includes three parts \cite{SM, Das2019LMC_model}:
\begin{equation}
\sigma^{\text{LMT}}_{\alpha\beta\gamma} = -\chi_{\alpha\beta\gamma} + \chi_\alpha \delta_{\beta\gamma} + \chi_\beta \delta_{\alpha\gamma}, \label{sigma_LMC}
\end{equation}
where the rank-3 tensor $\chi_{\alpha\beta\gamma}$ is given by
\begin{equation}
\chi_{\alpha\beta\gamma} = \frac{e^3\tau}{\hbar} \sum_n \int \frac{d^3 k}{(2\pi)^3}~ v^\alpha_{n\bm{k}} v^\beta_{n\bm{k}} \Omega^\gamma_{n\bm{k}} \left( -\frac{\partial f^0_{n\bm{k}}}{\partial\varepsilon_{n\bm{k}}} \right), \label{chi_abc}
\end{equation}
and the rank-1 tensors are given by $\chi_\alpha = \chi_{\alpha\gamma\gamma}$ and $\chi_\beta = \chi_{\beta\gamma\gamma}$. Einstein's convention of summation over repeated indices is assumed. In Eq. \eqref{chi_abc}, $f^0_{n\bm{k}}$ is the Fermi-Dirac distribution, $\tau$ is the relaxation time, $\bm{v}_{n\bm{k}}$ is the group velocity, and $\bm{\Omega}_{n\bm{k}}$ is the Berry curvature of the $n$-th band. The first term on the right-hand side of Eq. \eqref{sigma_LMC} (the rank-3 tensor $\chi_{\alpha\beta\gamma}$) originates from modified DOS while the following two (the rank-1 tensors $\chi_{\alpha,\beta}$) originate from chiral anomaly and $\bm{B}$-induced anomalous velocity as detailedly discussed in section S1 of SM \cite{SM}. Because $-\partial f^0_{n\bm{k}}/\partial \varepsilon_{n\bm{k}}\approx \delta(\mu - \varepsilon_{n\bm{k}})$ at low temperature with $\mu$ the chemical potential, the LMT is mainly determined by the electronic states near the Fermi surface. Due to the invariance of $\chi_{\alpha\beta\gamma}$ by exchanging $\alpha$ and $\beta$, LMT conductivity $\sigma^{\text{LMT}}_{\alpha\beta\gamma} B_{\gamma}$ is a symmetric tensor, distinguishing itself from the antisymmetric linear Hall conductivity $\sigma^{\text{H}}_{\alpha\beta\gamma}B_{\gamma}$ \cite{Tan2021UAHE}.

\begin{table}[htp]
\centering
\caption{List of all 122 magnetic point groups classified by the existence of rank-3 and rank-1 tensors $\chi_{\alpha\beta\gamma}$ and $\chi_{\alpha, \beta}$. $\checkmark$ means at least one tensor component is nonzero while $\times$ means the tensor is completely forbidden. In Berry-curvature-induced LMT, $\chi_{\alpha\beta\gamma}$ corresponds to the mechanism of modified density of states, and $\chi_{\alpha, \beta}$ correspond to mechanisms of chiral anomaly and $\bm{B}$-induced anomalous velocity.}\label{classification}
\begin{tabular}{p{7cm} c c}\\
\hline\hline
~~~~~~~~~~~~~~~Magnetic point groups & $\chi_{\alpha\beta\gamma}$ & $\chi_{\alpha, \beta}$ \\
\hline\\
11$^{\prime}$, $\overline{1}1^{\prime}$, $\overline{1}^{\prime}$, 21$^{\prime}$, m1$^{\prime}$, 2/m1$^{\prime}$, 2$^{\prime}$/m, 2/m$^{\prime}$, 2221$^{\prime}$, mm21$^{\prime}$, mmm1$^{\prime}$, m$^{\prime}$mm, m$^{\prime}$m$^{\prime}$m$^{\prime}$, 41$^{\prime}$, $\overline{4}1^{\prime}$, 4/m1$^{\prime}$, 4/m$^{\prime}$, 4$^{\prime}$/m$^{\prime}$, 4221$^{\prime}$, 4mm1$^{\prime}$, $\overline{4}2m1^{\prime}$, 4/mmm1$^{\prime}$, 4/m$^{\prime}$mm, 4$^{\prime}$/m$^{\prime}$m$^{\prime}$m, 4/m$^{\prime}$m$^{\prime}$m$^{\prime}$, 31$^{\prime}$, $\overline{3}1^{\prime}$, $\overline{3}^{\prime}$, 321$^{\prime}$, 3m1$^{\prime}$, $\overline{3}m1^{\prime}$, $\overline{3}^{\prime}$m, $\overline{3}^{\prime}$m$^{\prime}$, 61$^{\prime}$, $\overline{6}1^{\prime}$, 6/m1$^{\prime}$, 6$^{\prime}$/m, 6/m$^{\prime}$, 6221$^{\prime}$, 6mm1$^{\prime}$, $\overline{6}$m$21^{\prime}$, 6/mmm1$^{\prime}$, 6/m$^{\prime}$mm, 6$^{\prime}$/mmm$^{\prime}$, 6/m$^{\prime}$m$^{\prime}$m$^{\prime}$, 231$^{\prime}$, m$\overline{3}1^{\prime}$, m$^{\prime}\overline{3}^{\prime}$, 432, 4321$^{\prime}$, $\overline{4}3$m, $\overline{4}3$m1$^{\prime}$, m$\overline{3}$m, m$\overline{3}$m1$^{\prime}$, m$^{\prime}\overline{3}^{\prime}$m, m$^{\prime}\overline{3}^{\prime}$m$^{\prime}$ & $\times$ & $\times$ \\
\\ 222, mm2, mmm, 4$^{\prime}$, $\overline{4}^{\prime}$, 4$^{\prime}$/m, 422, 4$^{\prime}$22$^{\prime}$, 4mm, 4$^{\prime}$m$^{\prime}$m, $\overline{4}2$m, $\overline{4}^{\prime}$2$^{\prime}$m, $\overline{4}^{\prime}$2m$^{\prime}$, 4/mmm, 4$^{\prime}$/mm$^{\prime}$m, 32, 3m, $\overline{3}$m, 6$^{\prime}$, $\overline{6}^{\prime}$, 6$^{\prime}$/m$^{\prime}$, 622, 6$^{\prime}$22$^{\prime}$, 6mm, 6$^{\prime}$mm$^{\prime}$, $\overline{6}$m2, $\overline{6}^{\prime}$m$^{\prime}$2, $\overline{6}^{\prime}$m2$^{\prime}$, 6/mmm, 6$^{\prime}$/m$^{\prime}$mm$^{\prime}$, 23, m$\overline{3}$, 4$^{\prime}$32$^{\prime}$, $\overline{4}^{\prime}$3m$^{\prime}$, m$\overline{3}$m$^{\prime}$ & $\checkmark$ & $\times$ \\
\\ 1, $\overline{1}$, 2, 2$^{\prime}$, m, m$^{\prime}$, 2/m, 2$^{\prime}$/m$^{\prime}$, 2$^{\prime}$2$^{\prime}$2, m$^{\prime}$m2$^{\prime}$, m$^{\prime}$m$^{\prime}$2, m$^{\prime}$m$^{\prime}$m, 4, $\overline{4}$, 4/m, 42$^{\prime}$2$^{\prime}$, 4m$^{\prime}$m$^{\prime}$, $\overline{4}2^{\prime}$m$^{\prime}$, 4/mm$^{\prime}$m$^{\prime}$, 3, $\overline{3}$, 32$^{\prime}$, 3m$^{\prime}$, $\overline{3}$m$^{\prime}$, 6, $\overline{6}$, 6/m, 62$^{\prime}$2$^{\prime}$, 6m$^{\prime}$m$^{\prime}$, $\overline{6}$m$^{\prime}$2$^{\prime}$, 6/mm$^{\prime}$m$^{\prime}$ & $\checkmark$ & $\checkmark$\\ \\
\hline\hline
\end{tabular}
\end{table}

\textit{General symmetry properties of LMT.}---Because both $\bm{v}_{n\bm{k}}$ and $\bm{\Omega}_{n\bm{k}}$ are odd under time reversal symmetry (TRS, $\mathcal{T}$), tensors $\chi_{\alpha\beta\gamma}$ and $\chi_{\alpha,\beta}$ are forbidden by TRS. Besides, crystalline symmetries also place strong constraints on LMT (see details in section S2 of SM \cite{SM}). We further find that both $\chi_{\alpha\beta\gamma}$ and $\chi_{\alpha,\beta}$ must also vanish in those crystals with combined inversion-time-reversal symmetry $\mathcal{PT}$, and cubic point groups $O$, $T_d$, and $O_h$ including four-fold rotation symmetries $\{C_{4x}, C_{4y}, C_{4z}\}$ or improper rotation symmetries $\{\mathcal{P}C_{4x}, \mathcal{P}C_{4y}, \mathcal{P}C_{4z}\}$ even if TRS is broken. Moreover, $\chi_{\alpha,\beta}$ must vanish while $\chi_{\alpha\beta\gamma}$ survives in those crystals having $C_{nv}$ symmetry, combined rotation-time-reversal symmetry $C_n\mathcal{T}$ ($n\neq 2$), or two perpendicular rotation symmetries $C_n$ and $C'_n$. Then all the MPGs supporting Berry-curvature-induced LMT can be identified as summarized in Table \ref{classification}. It is found that 66 MPGs support $\chi_{\alpha\beta\gamma}$ among which 31 also support $\chi_{\alpha,\beta}$. Table \ref{classification} is a guiding principle to search for materials supporting Berry-curvature-induced LMT.

\begin{figure}[htp]
\includegraphics[width=\columnwidth]{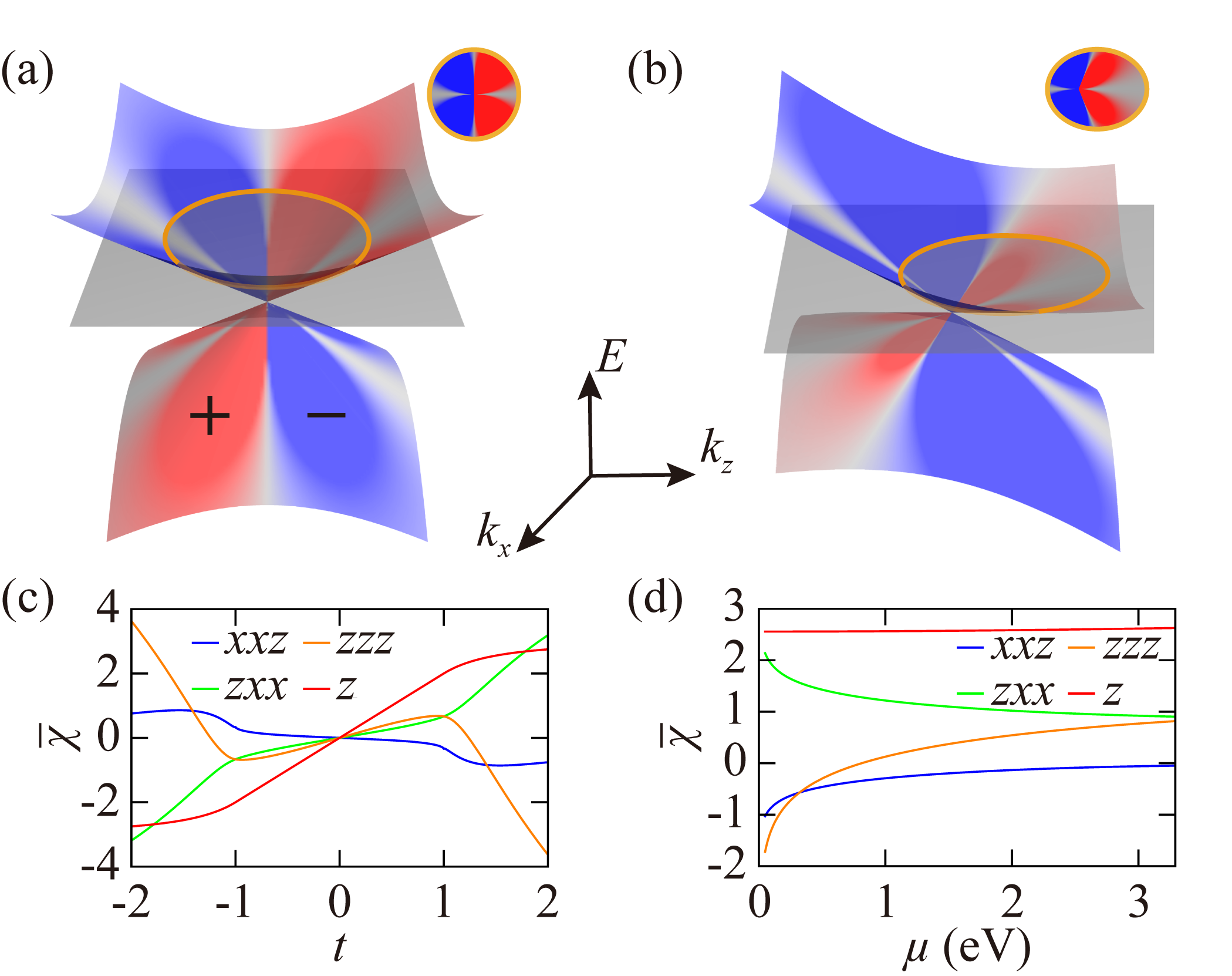}
\caption{(a)-(b) Schematics of band-resolved LMT tensor $\chi_{\alpha\beta\gamma}$ for (a) untilted and (b) tilted Weyl cones, respectively. Red (Blue) color represents the plus (minus) sign of $\chi_{zxx, n\bm{k}}{\sim}v_{n\bm{k}}^{z}v_{n\bm{k}}^{x}{\Omega}_{n\bm{k}}^{x}$ and the thickness of color represents its magnitude. Orange circles represent Fermi surfaces with their top view shown by insets. (c)-(d) Calculated LMT tensors of the tilted Weyl cone model [i.e., Eq. \eqref{twobandmodel}] with $\hbar v_{\rm F}=6.6$ eV$\cdot \mathrm{\AA}$. $\bar{\chi}$ is dimensionless, defined as $\chi/\chi_0$ with $\chi_0 = e^3\tau v_{\rm F}/2h^2$. (c) LMT tensors as functions of the tilting parameter $t=v_{\rm t}/v_{\rm F}$ with chemical potential $\mu=0.1$ eV. (d) LMT tensors as functions of chemical potential for a type-II Weyl cone with $t=1.5$.}\label{fig2}
\end{figure}

\begin{figure*}[htp]
\includegraphics[width=2\columnwidth]{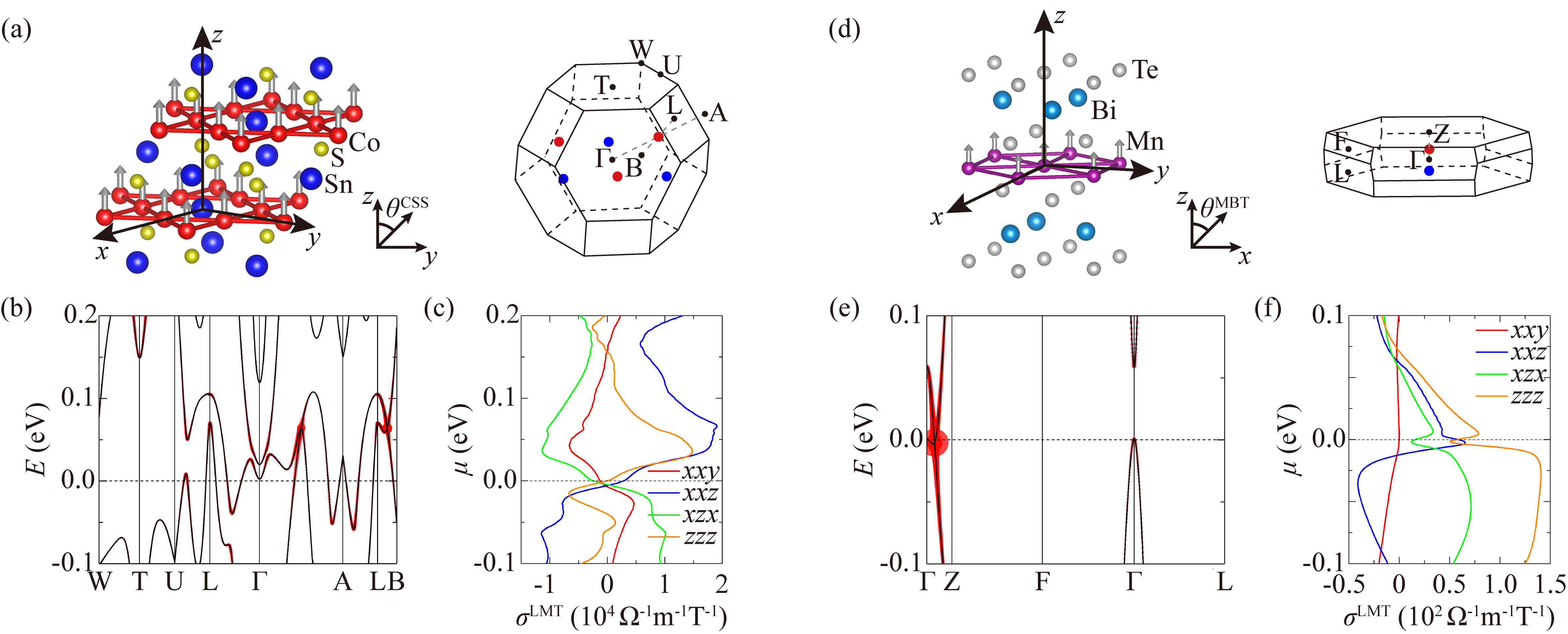}
\caption{Berry-curvature-induced LMT conductivities in magnetic Weyl semimetals (a)-(c) Co$_3$Sn$_2$S$_2$ and (d)-(f) FM MnBi$_2$Te$_4$. (a)(d) Crystal structure (left panel) and distribution of Weyl points (right panel) for Co$_3$Sn$_2$S$_2$ and FM MnBi$_2$Te$_4$. Gray arrows represent magnetic moments, and red (blue) dots within Brillouin zone represent Weyl points with positive (negative) chirality. $\theta^{\rm CSS}$ and $\theta^{\rm MBT}$ are angles between magnetization and $z$ axis, used in the following. $\theta^{\text{CSS}} = \theta^{\text{MBT}}=0$ here. (b)(e) Band structures including spin-orbit coupling. Size of the red dots indicates the magnitude of Berry curvature $|\bs{\Omega}_{n\bs{k}}|$. (c)(f) Independent nonzero components of $\sigma^{\text{LMT}}_{\alpha\beta\gamma}$ as functions of chemical potential $\mu$.}\label{fig3}
\end{figure*}

\textit{LMT in tilted Weyl cone model.}---The fact that Berry-curvature-induced LMT requires TRS breaking further motivates us to study magnetic WSMs. We start from considering a Weyl cone model with TRS broken by finite tilting:
\begin{equation}\label{twobandmodel}
    H_{\bm{k}} = s \hbar v_{\rm F} \bm{k} \cdot \bm{\sigma} + s\hbar v_{\rm t} k_z,
\end{equation}
where $s=\pm$ refers to the chirality of Weyl point (WP), $v_{\text{F}}$ is the Fermi velocity, and $t=v_{\rm t} /v_{\rm F}$ is a parameter describing the tilting of Weyl cone which is assumed along the $z$ direction. The Weyl cone model which is isotropic in the $k_xk_y$ plane allows only 4 independent nonzero LMT tensor components: $\chi_{xxz}$, $\chi_{zxx}$, $\chi_{zzz}$ and $\chi_z$, whose analytical expressions are calculated explicitly in section S3 of SM \cite{SM}. Their tilting dependence is shown in Fig. \ref{fig2}(c). For an untilted Weyl cone with $t=0$, LMT tensors vanish because of the cancellation of $v_{n\bm{k}}^{\alpha}v_{n\bm{k}}^{\beta}{\Omega}_{n\bm{k}}^{\gamma}$ between $\pm\bm{k}$ when integrating over the Fermi surface [Fig. \ref{fig2}(a)]. After increasing tilting breaks such a cancellation [Fig. \ref{fig2}(b)], LMT tensors exhibit an overall increase and reverse sign between $\pm t$ for both type-I ($|t|<1$) and type-II ($|t|>1$) Weyl cones. On the other hand, LMT tensors are independent of chemical potential $\mu$ for a type-I Weyl cone and insensitive to $\mu$ for a type-II Weyl cone as shown in Fig. \ref{fig2}(d). This is because the Berry curvature is proportional to $1/\mu^2$ while the Fermi surface area is proportional to $\mu^2$. Figure \ref{fig2}(d) only shows the $\mu>0$ region because LMT tensors are invariant between $\mu$ and $-\mu$. In magnetic WSMs, a pair of Weyl cones related by inversion symmetry contribute to LMT equally because of their simultaneously opposite tiltings and chiralities.

\textit{LMT in magnetic WSM materials.}---Then we investigate LMT concretely in magnetic WSMs Co$_3$Sn$_2$S$_2$ and FM MnBi$_2$Te$_4$ (see details of first-principles calculations in section S4 of SM \cite{SM}) whose structures and Brillouin zones are shown in Figs. \ref{fig3}(a) and (d), respectively. They have attracted much attentions because of their correlated magnetism and band topology \cite{xu2018topological,liu2018giant, Wang2018CSS_exp, morali2019fermi, liu2019magnetic,li2019intrinsic,li2019magnetically,zhang2019topological, gong2019experimental,liu2020robust,deng2020quantum}. They both belong to MPG $\overline{3}$m$^{\prime}$ including inversion $\mathcal{P}$, three-fold rotation $C_{3z}$ and the combined mirror-time-reversal $M_x \mathcal{T}$ symmetries, where $\chi_{\alpha\beta\gamma}$ and $\chi_{\alpha,\beta}$ both exist as shown in Table \ref{classification}.  Co$_3$Sn$_2$S$_2$ is one of the few experimentally confirmed magnetic WSMs with an out-of-plane FM ground state \cite{liu2018giant, Wang2018CSS_exp, morali2019fermi, liu2019magnetic}. There are totally three pairs of WPs related by $\mathcal{P}$ and $C_{3z}$ at about 60 meV above the Fermi level \cite{xu2018topological}. MnBi$_2$Te$_4$ possesses extraordinarily rich topological phases with an out-of-plane antiferromagnetic (AFM) ground state \cite{li2019intrinsic,li2019magnetically,zhang2019topological, gong2019experimental,liu2020robust,deng2020quantum} but can also be tuned to FM under a moderate magnetic field due to the weak interlayer AFM coupling \cite{zhu2020negative, li2021gate, huan2021magnetism, jiang2021quantum, lee2021evidence, lei2022magnetically}. FM MnBi$_2$Te$_4$ possesses a pair of $\mathcal{P}$-related WPs tilting along $z$ on -Z-$\Gamma$-Z near the Fermi level as predicted by calculations \cite{li2019intrinsic,li2019magnetically} and evidenced by experiments \cite{zhu2020negative, li2021gate, huan2021magnetism, jiang2021quantum, lee2021evidence, lei2022magnetically}.

Figure \ref{fig3}(b) gives the band structure and band-resolved Berry curvature distribution of Co$_3$Sn$_2$S$_2$ including spin-orbit coupling with a WP shown along $\Gamma$-A and L-B. Except for bands near WPs, considerable Berry curvature also distributes on bands near quasi-degenerate points, thus significantly contributing to LMT. Relaxation time of Co$_3$Sn$_2$S$_2$ is estimated as 1 ps by $\tau = m^{\ast}\mu_{\rm c} /e$, where effective mass $m^{\ast}$ and mobility $\mu_{\rm c}$ are adopted from experiments \cite{liu2018giant, ding2021quantum, ding2019intrinsic, shen2019anisotropies, zhou2020enhanced}. MPG $\overline{3}$m$^{\prime}$ allows four independent nonzero components of $\sigma_{\alpha\beta\gamma}^{\rm LMT}$ including $\sigma^{\rm LMT}_{xxy}$, $\sigma^{\rm LMT}_{xxz}$, $\sigma^{\rm LMT}_{xzx}$ and $\sigma^{\rm LMT}_{zzz}$ \cite{independentconducts}. Their chemical potential dependence is shown in Fig. \ref{fig3}(c), reaching an order of $10^4~{\rm\Omega^{-1}m^{-1}T^{-1}}$ and showing a sign reversal around $\mu=0$. Calculated LMT tensors in section S5 of SM \cite{SM} show that the only symmetry allowed rank-1 tensor component $\chi_{z}$ is much smaller than rank-3 tensor $\chi_{\alpha\beta\gamma}$ due to the cancellation between $\chi_{zxx}=\chi_{zyy}$ and $\chi_{zzz}$, indicating that Berry curvature induces LMT in Co$_3$Sn$_2$S$_2$ mainly by modifying DOS.

In FM MnBi$_2$Te$_4$, LMT is mainly induced by Berry curvature distributed around WPs as shown by band structure in Fig. \ref{fig3}(e). Relaxation time of FM MnBi$_2$Te$_4$ is estimated as 0.04 ps from experiments \cite{lee2021evidence, jiang2021quantum}. Figure \ref{fig3}(f) shows calculated $\sigma^{\rm LMT}_{\alpha\beta\gamma}$, reaching an order of $10^2~{\rm\Omega^{-1}m^{-1}T^{-1}}$. In addition the $\sigma^{\text{LMT}}_{xxy}$ is significantly smaller compared with other components, which is consistent with our previous model analysis where $\chi_{xxy}=-\sigma^{\rm LMT}_{xxy}$ vanishes due to the isotropy of ideal Weyl cone within the $k_xk_y$ plane. Small $\sigma^{\rm LMT}_{xxy}$ in FM MnBi$_2$Te$_4$ comes from weak anisotropy of WPs due to the three-fold rotation symmetry $C_{3z}$ along $z$ axis. According to the calculated LMT tensors in section S5 of SM \cite{SM}, mechanisms of modified DOS, chiral anomaly and $\bm{B}$-induced anomalous velocity all give significant contribution to LMT.

\begin{figure}
\includegraphics[width=\columnwidth]{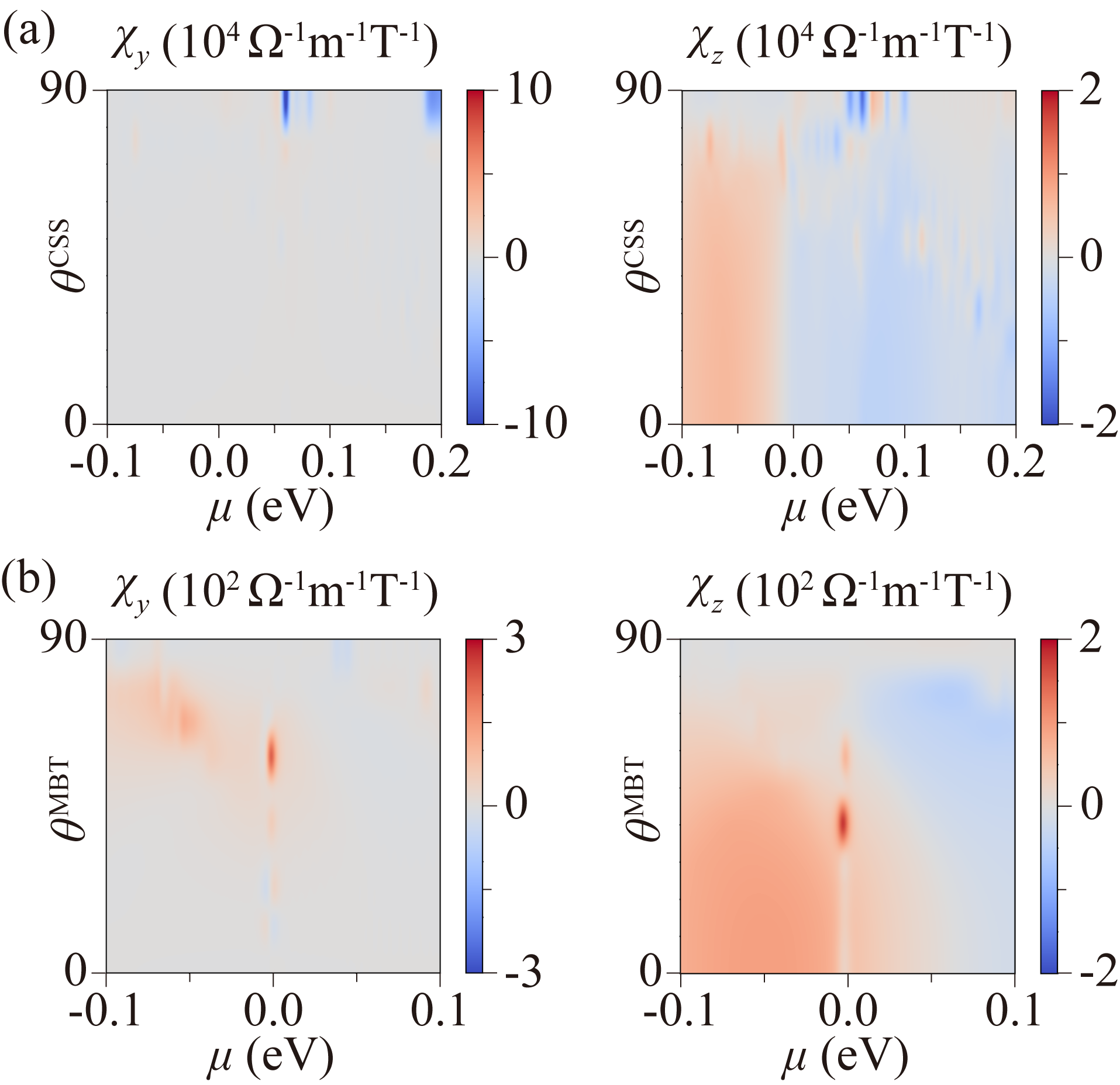}
\caption{Effects of magnetization canting (see the definition of angles $\theta^{\rm CSS}$ and $\theta^{\rm MBT}$ in the insets of Figs. \ref{fig3}(a) and (d), respectively) on Berry-curvature-induced LMT of (a) Co$_3$Sn$_2$S$_2$ and (b) FM MnBi$_2$Te$_4$. $\chi_{y,z}$ are taken as examples and other tensor components are shown in section S7 of SM \cite{SM}.}\label{fig4}
\end{figure}

\textit{Effects of magnetization canting.}---Although both Co$_3$Sn$_2$S$_2$ and FM MnBi$_2$Te$_4$ have easy axes along $z$, their magnetization directions can be canted away from the $z$ axis by external magnetic fields. In Co$_3$Sn$_2$S$_2$ and FM MnBi$_2$Te$_4$, required magnetic fields to overcome the magnetic anisotropy energy are less than 10 T and 2 T respectively (see section S6 of SM \cite{SM}), which are experimentally achievable. Figure \ref{fig4} shows calculated $\chi_y$ and $\chi_z$ of both materials as functions of canting angle and chemical potential. Other LMT tensor components exhibit similar qualitative features as shown in section S7 of SM \cite{SM}. Berry-curvature-induced LMT is generally robust against small magnetization canting introduced by perturbative magnetic fields, while is tunable by larger canting and greatly enhanced at certain canting angles. Our further analyses show that Berry-curvature-induced LMT is tuned by magnetization canting due to the coupling between Berry curvature and magnetization canting in Co$_3$Sn$_2$S$_2$ and also the change of band structure in FM MnBi$_2$Te$_4$ (see details in section S8 of SM \cite{SM}). The tunable LMT is promising for future experimental investigations.

\textit{Characteristic LMT signatures.}---Then we give all characteristic signatures to detect Berry-curvature-induced LMT experimentally. As aforementioned, Berry curvature only induces symmetric part of conductivity, i.e., the linear magnetoconductance (MC). While for resistivity, Berry curvature could induce both the symmetric linear MR and antisymmetric linear Hall resistivity, due to the mixture between Drude conductivity, anomalous Hall conductivity and linear MC. In section S9 of SM \cite{SM}, all conductivities and resistivities constrained by MPG $\overline{3}$m$^{\prime}$ are listed explicitly and discussed in detail.

\begin{figure}[htp]
\includegraphics[width=\columnwidth]{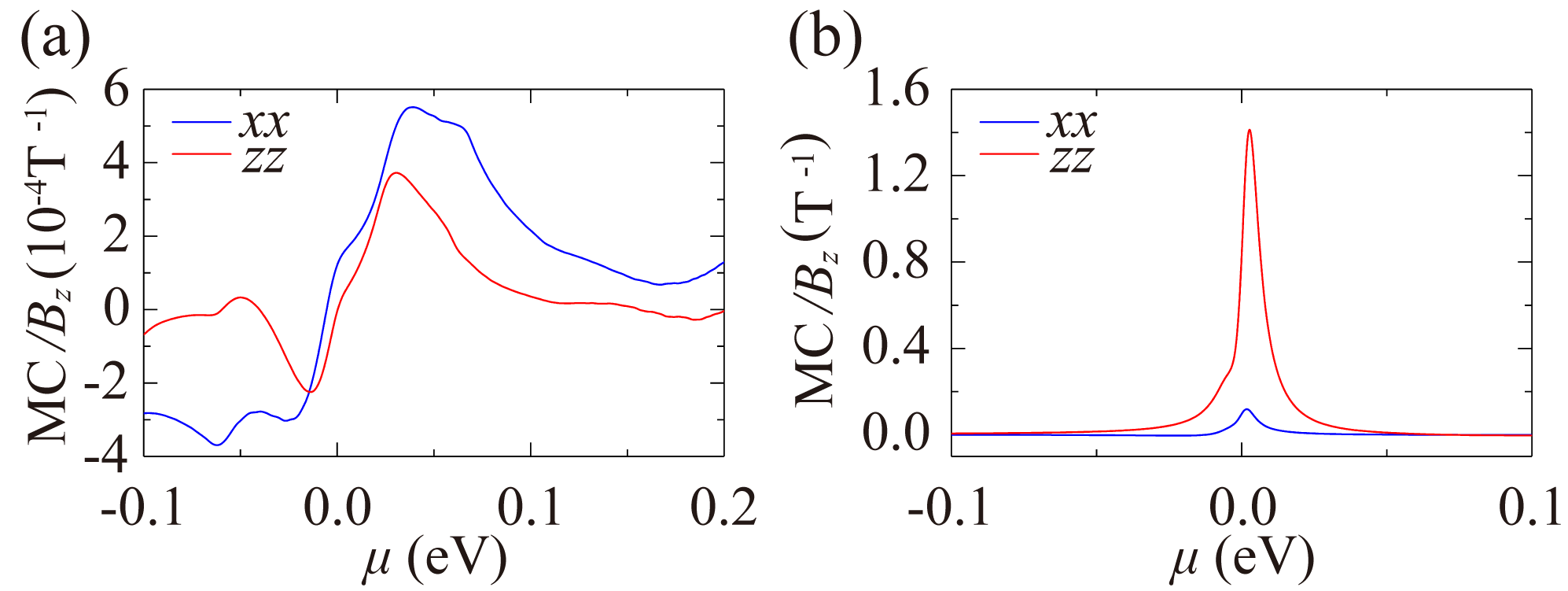}
\caption{Intrinsic linear magnetoconductance and magnetoresistance components $\mathrm{MC}_{\alpha\alpha}$ ($\alpha=x,y,z$) and $\mathrm{MR}_{zz}=\mathrm{MC}_{zz}$ of (a) Co$_3$Sn$_2$S$_2$ and (b) FM MnBi$_2$Te$_4$ when magnetic field and magnetization both point along $z$ axis. Because of $C_{3z}$ symmetry, $\mathrm{MC}_{xx} =\mathrm{MC}_{yy}$. }\label{fig5}
\end{figure}

Interestingly, we identify intrinsic components of linear MC and linear MR that are independent of relaxation time $\tau$, thus irrelevant to the quality of samples in experiments and solely determined by the intrinsic electronic structure. They serve as ideal quantities for comparison between theory and experiment. These components are
\begin{align}
\begin{split}
    \mathrm{MC}_{\alpha\alpha} &= \frac{\sigma_{\alpha\alpha}(\bs{B}) - \sigma_{\alpha\alpha}(0)}{\sigma_{\alpha\alpha}(0)} =\frac{\sigma^{\text{LMT}}_{\alpha\alpha\gamma}B_\gamma}{\sigma^{\rm Drude}_{\alpha\alpha}},
\end{split}\nonumber\\
\begin{split}
    \mathrm{MR}_{zz} &= \frac{\rho_{zz}(\bs{B}) - \rho_{zz}(0)}{\rho_{zz}(0)} =\frac{\sigma^{\text{LMT}}_{zz\gamma}B_\gamma}{\sigma^{\rm Drude}_{zz}},
\end{split}\label{MRaa}
\end{align}
where $\alpha=x,y,z$. Their dependence on $\tau$ is eliminated because Drude conductivity $\sigma^{\rm Drude}_{\alpha\alpha}$ and LMT conductivity $\sigma^{\text{LMT}}_{\alpha\beta\gamma}B_{\gamma}$ are both proportional to $\tau$. For MnBi$_2$Te$_4$ whose ground state is AFM, $\sigma_{\alpha\alpha}(0)$ and $\rho_{zz}(0)$ of FM MnBi$_2$Te$_4$ can be extrapolated from data at higher magnetic fields where FM state is stabilized. According to Eq. \eqref{MRaa}, materials with a large $\sigma^{\text{LMT}}_{\alpha\alpha\gamma}$ but a small $\sigma^{\rm Drude}_{\alpha\alpha}$, i.e., materials with large Berry curvature but small Fermi surface, such as magnetic WSMs, are favored by these intrinsic signatures. Figure \ref{fig5}(a) and (b) show the ${\rm MC}_{\alpha\alpha}$ and ${\rm MR}_{zz}$ for Co$_3$Sn$_2$S$_2$ and FM MnBi$_2$Te$_4$, respectively, exhibiting an enhancement when $\mu$ approaches the energy of WPs. Due to large Fermi surfaces contributed by bands irrelevant to WPs, they are small in Co$_3$Sn$_2$S$_2$. While for FM MnBi$_2$Te$_4$ with small Fermi surfaces, $\mathrm{MR}_{zz}$ and $\mathrm{MC}_{zz}$ reach a large value exceeding 100\% per tesla, which can be easily examined by future experiments.

\textit{Comparison with experiments.}---Finally, we consider recent LMT experiments \cite{jiang2021chirality,li2021gate, lei2020surface, zhu2020negative,huan2021magnetism}, which can not be explained by previous theories \cite{parish2003non, parish2005classical,abrikosov1998quantum, abrikosov2000quantum}. In Co$_3$Sn$_2$S$_2$, linear MR components ${\rm MR}_{yx}(B_x)$ and ${\rm MR}_{xx}(B_y)$ are observed \cite{jiang2021chirality}. Corresponding components of $\sigma^{\rm LMT}_{\alpha\beta\gamma}$ are $\sigma^{\rm LMT}_{yxx}{=}1360~{\rm \Omega^{-1}m^{-1}T^{-1}}$ and $\sigma^{\rm LMT}_{xxy}{=}1300~{\rm \Omega^{-1}m^{-1}T^{-1}}$ from experimental data \cite{methodfromexp}. In our calculations, $\sigma^{\rm LMT}_{yxx}=\sigma^{\rm LMT}_{xxy}$ reaches a maximum of $6600~{\rm \Omega^{-1}m^{-1}T^{-1}}$ when $\mu$ is near the energy of WPs and becomes several times smaller after $\mu$ shifts slightly away \cite{cantingeffect}, in agreement with the experiment. In FM MnBi$_2$Te$_4$ and MnSb$_2$Te$_4$ whose band structure is similar to that of MnBi$_2$Te$_4$, ${\rm MR}_{xx}(B_z)$ is observed \cite{li2021gate,lei2020surface, zhu2020negative,huan2021magnetism}. Corresponding $\sigma^{\rm LMT}_{xxz}$ in our calculations (about $60~{\rm\Omega^{-1} m^{-1}T^{-1}}$ when $\mu$ is near the energy of WPs) is similar to that in six-layer MnBi$_2$Te$_4$ (from $8$ to $200~{\rm \Omega^{-1}m^{-1}T^{-1}}$) but much smaller than that in thick MnBi$_2$Te$_4$ flakes and MnSb$_2$Te$_4$ bulk (about $2000~{\rm \Omega^{-1}m^{-1}T^{-1}}$). Such differences may originate from the variation of relaxation time $\tau$. ${\rm MC}_{xx}(B_z)$ insensitive to $\tau$ ranges from $0.23\%$ to $8\%$ per tesla from experiments, equaling to our results when $\mu$ is away from the energy of WPs by several or tens of meV [Fig. \ref{fig5}(b)]. Our results also agree with the observed dependence of linear MR on magnetic field direction and chemical potential (see details in section S10 of SM \cite{SM}). Moreover, from Eq. \eqref{sigma_LMC}, only the rank-3 tensor $\chi_{\alpha\beta\gamma}$ contributes to $\sigma^{\text{LMT}}_{\alpha\beta\gamma}$ components above because the relevant rank-1 tensor components $\chi_{y}$ vanish due to the $C_{3z}$ symmetry. Since our numerical calculations based purely on Berry curvature effect already agree well with currently available experimental data,  we conclude that Berry curvature effect dominates linear MR observed in these materials and the rank-3 tensor $\chi_{\alpha\beta\gamma}$ (i.e., the Berry-curvature-modified DOS) plays a major role.

\textit{Summary.}---In summary, the crucial role of Berry curvature in LMT of realistic magnetic materials is revealed. Berry-curvature-induced LMT is a general phenomenon existing in 66 out of 122 MPGs according to the presented symmetry guideline. Focusing on representative magnetic WSMs Co$_3$Sn$_2$S$_2$ and FM MnBi$_2$Te$_4$, the contribution of Berry curvature to LMT is estimated concretely, which is tunable by magnetization canting. Berry-curvature-induced LMT can be justified by various characteristic experimental signatures, especially an intrinsic MR exceeding 100\% per tesla. Our results suggest that observed LMT in recent experiments on Co$_3$Sn$_2$S$_2$ and FM MnBi$_2$Te$_4$ are actually dominated by Berry curvature through modifying DOS.

\begin{acknowledgments}
We acknowledge fruitful discussions with Prof. Binghai Yan and Prof. Yong Xu. This work was supported by the Basic Science Center Project of NSFC (Grant No. 51788104), the Ministry of Science and Technology of China, and the Beijing Advanced Innovation Center for Future Chip.
\end{acknowledgments}


\begin{thebibliography}{79}
\bibitem [{met()}]{methodfromexp}%
  \BibitemOpen
  \href@noop {} {}\bibinfo {note} {LMT resistivity, zero-magnetic-field
  resistivity (Drude resistivity) and anomalous Hall resistivity are directly
  available from experimental data. Then all corresponding quantities can be
  converted according to expressions of all resistivities and conductivities as
  shown in section S9 of SM \cite{SM}.}\BibitemShut {Stop}%
\bibitem [{can()}]{cantingeffect}%
  \BibitemOpen
  \href@noop {} {}\bibinfo {note} {Magnetization canting effects are ignored
  here since Berry-curvature-induced LMT is robust against limited
  magnetization canting as discussed above.}\BibitemShut {Stop}%
\end{thebibliography}
\end{document}